\documentclass[PRB,twocolumn,doublespace,showpacs,preprintnumbers,superscriptaddress]{revtex4}
\usepackage{amsmath}
\usepackage{dcolumn}
\usepackage{bm}
\usepackage{longtable}
\usepackage{graphicx}

\def \a {$\hat a$ }
\def \b {$\hat b$ }
\def \c {$\hat c$ }

\begin{document}

\title{Anisotropy in magnetic and transport properties of Fe$_{1-x}$Co$_{x}$%
Sb$_{2}$}
\author{Rongwei Hu}
\affiliation{Condensed Matter Physics and Materials Science Department, Brookhaven
National Laboratory, Upton, NY 11973}
\affiliation{Department of Physics, Brown University, Providence, RI 02912}
\author{V. F. Mitrovi{\'c}}
\affiliation{Department of Physics, Brown University, Providence, RI 02912}
\author{C. Petrovi{\'{c}}$^{1}$}
\date{\today}

\begin{abstract}
Anisotropic magnetic and electronic transport measurements were carried out
on large single crystals of Fe$_{1-x}$Co$_{x}$Sb$_{2}$ ($0\leq x\leq 1$).
The semiconducting state of FeSb$_{2}$ evolves into metallic and weakly
ferromagnetic by substitution of Fe with Co for $x<0.5$. Further doping
induces structural transformation from orthorhombic $Pnnm$ structure of FeSb$%
_{2}$ to monoclinic $P21/c$ structure of CoSb$_{2}$ where semiconducting and
diamagnetic ground state is restored again. Large magnetoresistance and
anisotropy in electronic transport were observed.
\end{abstract}

\maketitle

\section{Introduction}

Previous attempts to explain the unusual magnetic and electronic properties
of the narrow-gap semiconductor FeSi$^{1,2}$ were based on either the
itinerant magnetism model, the Kondo description, the mixed valence model,
or the two-band Hubbard model, and started with Jaccarino's interpretation
of an extreme narrow-band-energy-gap model.$^{3-6}$ From this work two
different viewpoints have emerged. The first, proposed by Takahashi and
Moriya,$^{3}$ treats FeSi as a nearly ferromagnetic semiconductor in an
itinerant electron model. The second, as pointed out by Aeppli and Fisk,$%
^{4} $ takes FeSi as a transition metal version of the Kondo insulator with
localized Fe moments. This viewpoint has the origin in the narrow gap/high
density of states model that Jaccarino considered and then rejected.
Experimental confirmation of that model was achieved by Mandrus et al$^{7}$
and Park et al.$^{8}$ Other analogous $3d$ model systems are highly desired
in order to study this problem, particularly those of non-cubic and strongly
anisotropic structures.

The magnetic properties of FeSb$_{2}$ strongly resemble those of FeSi.$%
^{9,10}$ Previous investigation of FeSb$_{2}$ has shown that the magnetism
can be described by two effects: a thermally induced Pauli susceptibility
and a low spin to high spin transition within the t$_{2g}$ multiplet of Fe
ion in octahedral crystal field.

In order to show the possible Kondo insulator to heavy-fermion metal
transition, we perturbed the FeSb$_{2}$ electronic system by electron
doping. We report the anisotropy in magnetic and electronic transport
properties of Fe$_{1-x}$Co$_{x}$Sb$_{2}$ as a function of Co content. The
semiconducting ground state of FeSb$_{2}$ evolves into a metallic one by $%
x=0.1$. Further Co substitution induces weak ferromagnetism for $0.2\leq
x<0.5$. Beyond $x=0.5$ there is a structural transformation from
orthorhombic $Pnnm$ to the monoclinic $P21/c$ structure of CoSb$_{2}$ and
the restoration of the anisotropic semiconducting ground state for $0.5\leq
x<1$. The weak ferromagnetism and high temperature Curie-Weiss behavior is
due to the small portion of localized Co ions. The large positive
magnetoresistance arises in the Fe$_{1-x}$Co$_{x}$Sb$_{2}$ electronic
system, particularly for $x$ values close to the metal - semiconductor
crossover.

\vspace*{-0.4cm}

\section{Experimental Details}

Single crystals were grown with the high temperature flux method.$^{11,12}$
Their structure and composition were determined by analyzing powder X-ray
diffraction (XRD) spectra taken with Cu K$\alpha $ radiation ($\lambda
=1.5418$ \AA ) using a Rigaku Miniflex X-ray machine. The lattice parameters
were obtained by fitting the XRD spectra using the Rietica software.$^{13}$
A JEOL JSM-6500 SEM microprobe was used for verifying the Co concentration.
Crystals were oriented using a Laue Camera and polished into rectangular
bars along specific crystalline axes. Electrical contacts were made with
Epotek H20E silver epoxy for a standard 4-wire resistance measurement. The
dimensions of the samples were measured by a high precision optical
microscope with 10 $\mu \mathrm{m}$ resolution. Magnetization, resistivity,
and transverse magnetoresistance measurements were carried out in a Quantum
Design MPMS-5 and a PPMS-9 varying the temperature from 1.8 K to 350 K and
applying a magnetic field up to 90 kOe. Polycrystalline averages were
calculated from $M/H\equiv \chi =1/3\,[\chi _{a}+\chi _{b}+\chi _{c}]$. Heat
capacity data was collected in the PPMS-9 instrument using a relaxation
technique. \vspace*{-0.2cm}

\section{Results and Discussion}

\subsection{Crystal Structure}

Fig. 1 shows the lattice constants and unit cell volume of Fe$_{1-x}$Co$_{x}$%
Sb$_{2}$ as a function of nominal Co content, as determined by powder X-ray
diffraction. As can be seen from Fig. 1, the end member structures of Fe$%
_{1-x}$Co$_{x}$Sb$_{2}$ are orthorhombic $Pnnm$ and monoclinic $P21/c$ for
FeSb$_{2}$ and CoSb$_{2}$, respectively. Both \a and \b{a}xes show a smooth
linear decrease with increasing $x$ for $0\leq x<0.5$. In contrast, the \c{a}%
xis expands with increasing $x$ in the same doping range. At $x=0.5$, the \c{%
a}xis doubles in size and a monoclinic angle $\beta \neq 90{{}^{\circ }}$
emerges, indicating a structural phase transition to lower symmetry of the
unit cell.

In the monoclinic region, for $0.5\leq x<1$, the unit cell volume refers to
that of a pseudo-marcasite cell. This cell is related to the true unit cell
of CoSb$_{2}$ (specified by $\overrightarrow{a}^{\prime },\overrightarrow{%
\text{ }b}^{\prime },\overrightarrow{c}^{\prime },\,\mathrm{and}\,\beta
^{\prime })$ by vectorial equations of the following form: $\overrightarrow{a%
}^{\prime }=(\overrightarrow{a}-\overrightarrow{c})/2,\overrightarrow{\text{ 
}b}^{\prime }=\overrightarrow{b}$ and $\overrightarrow{c}^{\prime }=(%
\overrightarrow{a}+\overrightarrow{c})/2$.$^{14}$ In this doping regime as
well, we observe a smooth evolution of a, b and c lattice parameters with
increasing $x$. The evolution is concurrent with a gradual increase of the
monoclinic angle $\beta ^{\prime }$, from 90.1${{}^{\circ }}$ to 90.4${%
{}^{\circ }}$ (Fig. 1(Inset)). This linear dependence on Co concentration,
conforming to Vegard's law, demonstrates that Co successfully substitutes
for Fe in the entire doping range. In addition, energy dispersive SEM for
several nominal x=0.25 samples showed that the uncertainty in Co
concentration among samples grown from different batches was $\Delta x=0.04$.

The anisotropy in the evolution of lattice parameters with Co content for $%
0\leq x<0.5$ is consistent with an increased occupancy of $d_{xy}$ orbitals,
directed towards nearest neighbor Fe ions along \c{a}xis of the crystal.$%
^{10}$ Monoclinic distortion at $x=0.5$ increases the \c{a}xis and decreases
the overlap of d$_{xy}$ orbitals. The reduced overlap results in an
enlargement of the gap within the t$_{2g}$ multiplet$^{15}$ and causes a
change in electronic transport properties. The electronic system gradually
evolves towards a semiconducting ground state of CoSb$_{2}$ with the
increase of Co concentration from $0.5\leq x<1$. 
\begin{figure}[t]
\centerline{\includegraphics[scale=0.65]{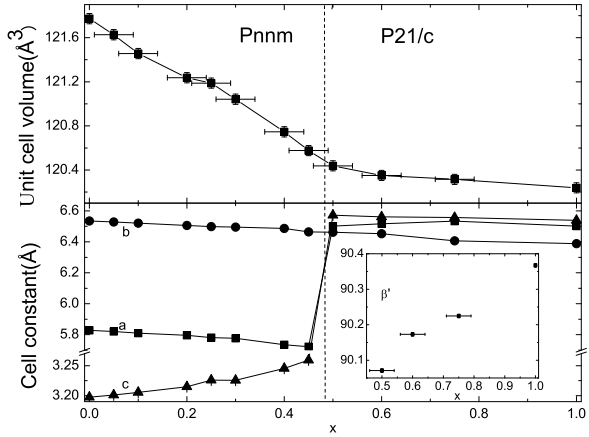}} 
\vspace*{-0.2cm}
\caption{{\protect\small Unit cell constants and volume of the series Fe$%
_{1-x}$Co$_{x}$Sb$_{2}$ as a function of composition. Phase transition takes
place at $x=0.5$. Dashed line represents the phase boundary between
orthorhombic and monoclinic structures. Inset shows monoclinic angle $%
\protect\beta ^{\prime }$ in P21/c structure.} }
\label{Fig1}
\end{figure}
%
%
\vspace*{-0.3cm} %
%
%
%
%
\begin{figure}[tbph]
\centerline{\includegraphics[scale=0.5]{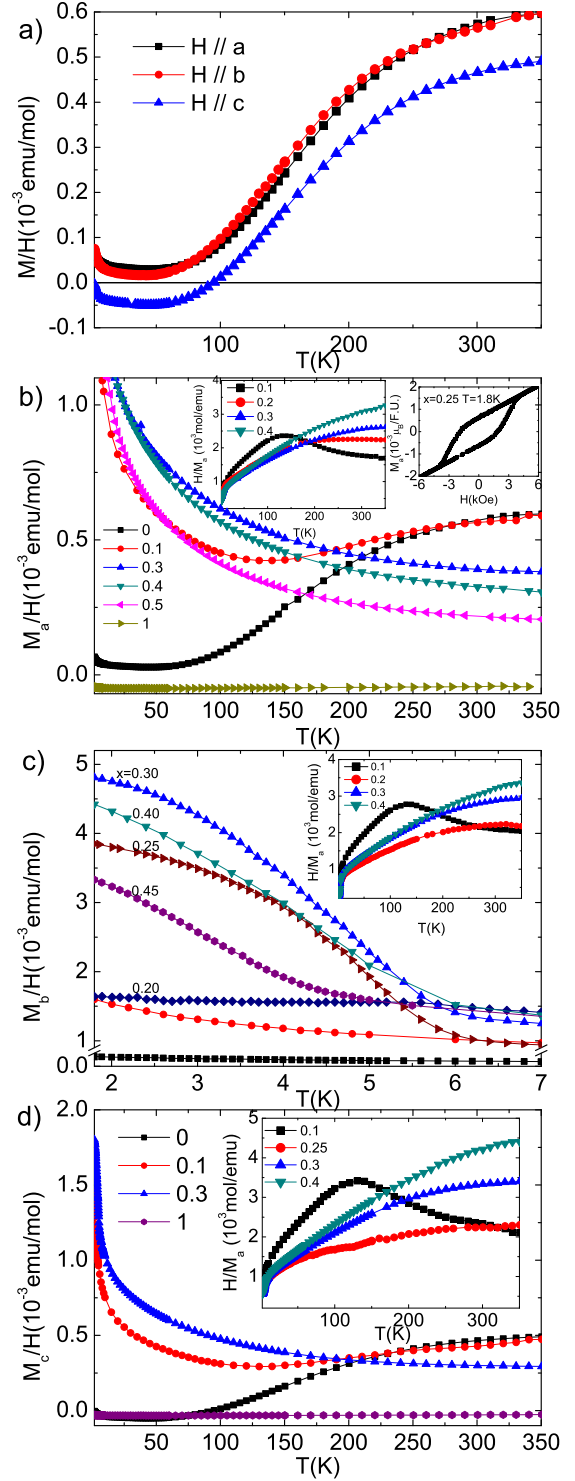}} 
\vspace*{-0.5cm}
\caption{{\protect\small $M/H\,\mathrm{v.s.}\,T$ at $H=1000$ Oe along 3
crystalline axes. Some data of the series were omitted for clarity: \textbf{%
\ (a)} magnetization of FeSb$_{2}$ along all three axes. \textbf{\ (b)}%
Magnetization along \a{a}xis. Insets show the 1/}$\protect\chi $%
{\protect\small \ and also the hysteresis loop of Fe$_{0.75}$Co$_{0.25}$Sb$%
_{2}$. \textbf{(c)} Magnetization along \b{a}xis displays weak ferromagnetic
transition at low temperatures for $0.2\leq x\leq 0.45$. \textbf{(d)} \c{a}%
xis magnetization resembles those of \a axis but shows no transitions at low
temperatures. }}
\label{Fig2}
\end{figure}

%

\subsection{Magnetic Properties}


\bigskip 

\begin{table*}[t]
\caption{Parameters of the fits to the polycrystalline average of $M/H$ data
for $0\leq x<0.5$.}%
\begin{tabular}{cccccccccc}
\hline\hline
\multicolumn{10}{c}{$\chi (T)=\chi _{NB}+\chi _{CW}$} \\ \hline
$x$ & $\Delta _{\chi }(K)$ & $W(K)$ & $p$ & $\chi _{0}(10^{-4}emu/mol)$ & $%
\mu _{eff}(\mu _{B})$ & $\Theta (K)$ & $\mu _{eff}/Co(\mu _{B})$ & $T_{c}(K)$
& $Ps(10^{-3}\mu _{B})$ \\ \hline
$0$ & 
\begin{tabular}{l}
$552$ \\ 
$425$%
\end{tabular}
& 
\begin{tabular}{l}
$1.6$ \\ 
$310$%
\end{tabular}
& 
\begin{tabular}{l}
$1.3$ \\ 
$1.4$%
\end{tabular}
& 
\begin{tabular}{l}
$0.03$ \\ 
$0.04$%
\end{tabular}
& $0$ & $0$ &  &  &  \\ 
$0.05$ & $531$ & $235$ & $2.1$ & $0.3$ & $0.41$ & $-23$ &  &  &  \\ 
$0.10$ & $511$ & $471$ & $2.5$ & $1.6$ & $0.79$ & $-71$ &  &  &  \\ 
$0.20$ &  &  &  & $4.5$ & $0.14$ & $110$ & $0.70$ & $6.6$ & $1.49$ \\ 
$0.25$ &  &  &  & $3.3$ & $0.24$ & $104$ & $0.99$ & $6.0$ & $1.73$ \\ 
$0.30$ &  &  &  & $3.0$ & $0.32$ & $77$ & $1.08$ & $6.8$ & $1.87$ \\ 
$0.40$ &  &  &  & $1.5$ & $0.54$ & $17$ & $1.35$ & $7.3$ & $1.19$ \\ 
$0.45$ &  &  &  & $1.5$ & $0.54$ & $16$ & $1.20$ & $6.4$ & $0.75$ \\ \hline
\end{tabular}%
\label{Tb1}
\end{table*}

For $x=0$, the temperature dependence of the susceptibility is qualitatively
the same for $H\,||\,\hat{a},\hat{b},\,\mathrm{and}\,\ \hat{c}$ as shown in
Fig. 2(a). A diamagnetic to paramagnetic crossover at $T$ in the vicinity of
100 K occurs when a field is applied parallel to the $\hat{c}$ axis. This
crossover is absent when fields are applied parallel to the $\hat{a}$ and $%
\hat{b}$ axes. Furthermore, for $H\,||\,\hat{a},\hat{b}$ below 100 K the
susceptibility is temperature independent and can be described by a
paramagnetic term $\chi _{a}=\chi _{b}=2\times 10^{-5}$ $emu/mol$. These
observations contradict previous measurements of FeSb$_{2}$,$^{9}$ that
reported a diamagnetic to paramagnetic crossover in the vicinity of 100 K
for fields applied along all three axes. Our measurements were performed on
large single crystal samples which eliminated the need for subtracting the
background, which is often difficult to estimate with high certainty.
Subtraction of the background was necessary in Ref. 9 due to very small mass
of the FeSb$_{2}$ samples. Consequently, the intrinsic susceptibility of FeSb%
$_{2}$ determined on large single crystals, differs from Refs. 9 and 10 and
is shown in Fig.2(a).

In what follows, we explore to what extent different models can account for
the observed paramagnetic moment and the temperature dependence of the
susceptibility in Fe$_{1-x}$Co$_{x}$Sb$_{2}$. In the free-ion model the
susceptibility $\chi $ is given by Jaccarino et al$^{2}$ as, 
\begin{equation*}
\chi _{FI}=Ng^{2}\mu _{B}^{2}\frac{J(J+1)}{3k_{B}T}\frac{2J+1}{2J+1+\exp
(\Delta _{\chi }/k_{B}T)}+\chi _{0}.
\end{equation*}%
This model describes thermal excitations from low to high spin states,
separated by a spin gap $\Delta _{\chi }$. Moreover in the simplest case,
two spin states can be viewed as the $J=S=0$ spin singlet and the $S=1$
triplet state. Fitting the expression to the polycrystalline average of the
magnetic susceptibility of FeSb$_{2}$ and setting $J=S$, we obtain $\chi
_{0}=4.5\times 10^{-6}emu/mol$, $\Delta _{\chi }=527$ $K$ and $S=0.24$. The
rather small value of $S$ implies that the free-ion model may not be the
explanation for the thermally induced paramagnetism at high temperatures, or
that a small portion of Fe spins in the S=0 low spin d$^{4}$ ground state(t$%
_{2g}$,S=0) takes part in the spin-state transition.

The magnetic susceptibility of FeSb$_{2}$ can also be described by the
narrow-band-small-gap picture:

\begin{equation*}
\chi _{NB}=\frac{2N\mu_{B}^{2}p \exp (\Delta _{\chi }/T)(\exp (W/T)-1)}{%
W(1+\exp (\Delta _{\chi }/T))(1+\exp ((\Delta _{\chi }+W)/T))}+\chi _{0}
\end{equation*}
\vspace*{-0.1cm}

Here $\chi _{NB}$ is the Pauli susceptibility of an itinerant electron
system with a density of states $N(E)$ consisting of two narrow bands of
width $W$ separated by a gap $E_{g}\equiv 2\Delta _{\chi }$, so that the
chemical potential is defined as $\mu =W+\Delta _{\chi }$. The density of
states is given by $N(E)=Np/W$, where $N$ denotes the number of unit cells
and $p$ the number of states per unit cell. In this picture, $\chi _{0}$ is
the temperature independent paramagnetic susceptibility. By fitting this
model, we find the allowable values of parameters: $1.3<p<1.4,$ $%
1.6K<W<310K, $ $850K<E_{g}<1100K$. The result agrees with Ref.9 but differs
in a positive Pauli term, $\chi _{0}=3\times 10^{-6}-4\times 10^{-6}$
emu/mol, instead of a diamagnetic term.

%
%
%
\begin{figure}[t]
\centerline{\includegraphics[scale=0.56]{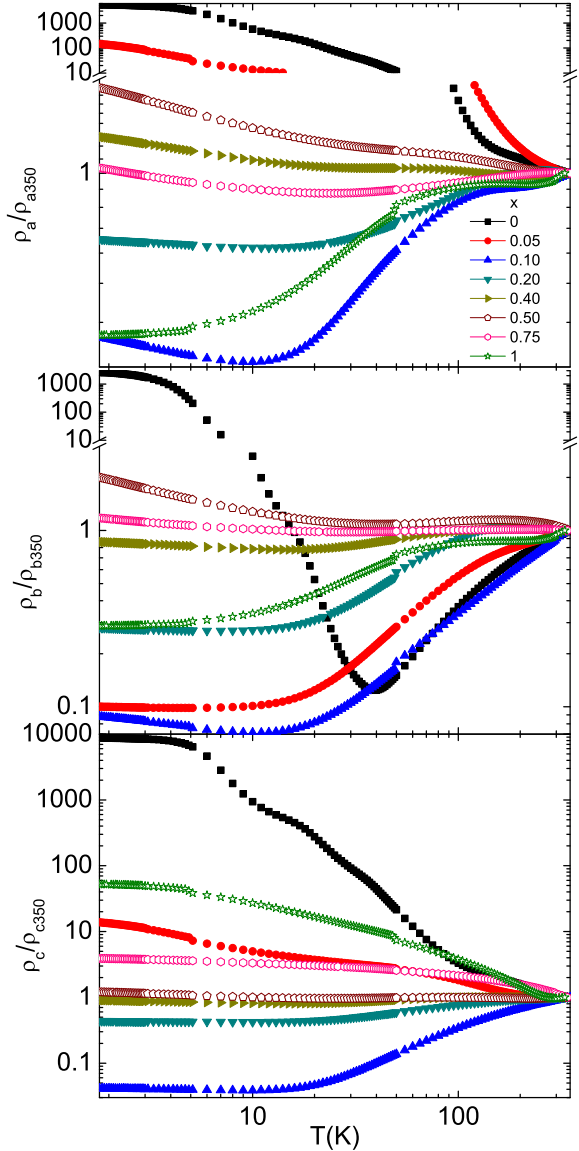}} 
\vspace*{-0.2cm}
\caption{{\protect\small Normalized resistivity along all three axes as a
function of temperature and different doping levels, $x$ denoted by
different shapes and colors. Fe$_{1-x}$Co$_{x}$Sb$_{2}$ becomes metallic
along all axes for $x>0.05$. With increasing $x$, the metallicity is lost in
a crossover, where resistivity is nearly independent of temperature. Further
increase of $T$ leads to metallic behavior again along $\hat{a}$ and $\hat{b}
$. However, resistivity along $\hat{c}$ axis evolves gradually from metallic
to semiconducting. Below 40 K for $x=0$ and 0.05 samples the impurity tails
up to 5 $\Omega $cm are observed. }}
\label{Fig3}
\end{figure}
%
%
At high temperatures for $0< x\leq 0.45$, the magnetic susceptibility along
all three crystalline axes can be described by two terms: a thermally
induced paramagnetic moment with progressively smaller gap $\Delta _{\chi }$%
, and a growing Curie-Weiss term, due to Co substitution:

\begin{equation*}
\chi (T)=\chi _{CW}+\chi _{NB}
\end{equation*}%
where 
\begin{equation*}
\chi _{CW}=\frac{N\mu _{eff}^{2}}{3k_{B}(T-\Theta )}
\end{equation*}

Results of fits for $0\leq x<0.5$ are shown in Table I. The gap $\Delta
_{\chi }$ decreases with increasing doping as $x$ reaches 0.2. For $0.20\leq
x\leq 0.45$ the temperature dependence of the magnetic susceptibility above
180 K is characterized solely by the Curie-Weiss term and temperature
independent Pauli term $\chi _{0}$.

The low temperature magnetic properties of Fe$_{1-x}$Co$_{x}$Sb$_{2}$ are
anisotropic. For $H||\hat{a}\ \mathrm{and}\ \hat{b}$, we observe a large
increase of $M/H$ values, a signature of ferromagnetic transition.
Consequently, hysteresis loops, shown in the inset to Fig. 2 (b), are
observed for 0.2$\leq x\leq 0.45$ ($H||\hat{b}$), as well as for $x=0.25$ ($%
H||\hat{a}$) at T=1.8K. The saturation moments P$_{s}$ are listed in Table
I. The Curie temperatures, determined by extrapolating the steepest slope of
the M/H curve to $M=0$, vary from 5.9 to 7.2 K (Tab I.). This corresponds to
the anomalies in the heat capacity measurements (Fig. 6). CoSb$_{2}$ is
diamagnetic in the whole experimental temperature range. As Fe is in the
nonmagnetic 3d$^{4}$ configuration, we assume that localized moment arises
solely from the substituted Co atoms, and hence Curie-Weiss term. To
evaluate the high temperature moment per Co atom, we fit the polycrystalline
average of the magnetic susceptibility for temperatures ranging from 180 to
350 K with the Curie-Weiss law. We find the moment of $\mu /\mathrm{Co}$ to
be $1\pm 0.35\mu _{B}$ for all the ferromagnetic $x$ (Tab. I). This suggests the existence of polarized
moments of Co$^{2+}$ with low spin and quenched orbital momentum in the
lattice. Therefore, we conclude that the weak ferromagnetism is due to the
ordering of 0.2 \% of Co$^{2+}$ moments. The fact that we were unable to
detect any hysteresis loops above T$_{c}$ and the absence of any sample
dependence among sample within one batch of crystals, as well as among
crystals from different batches, argues against extrinsic sources of weak
ferromagnetism in Fe$_{1-x}$Co$_{x}$Sb$_{2}$.

%
\begin{figure}[t]
\centerline{\includegraphics[scale=0.58]{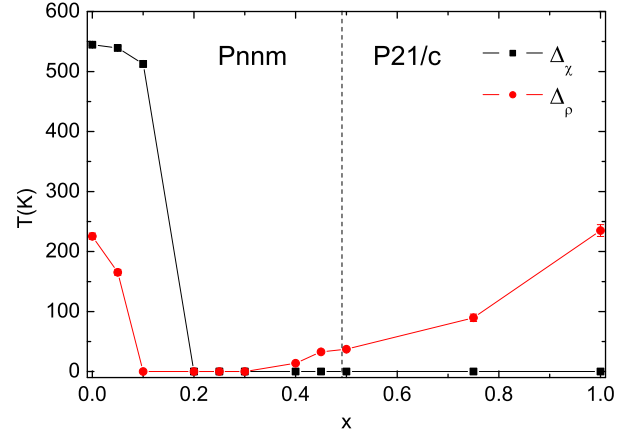}} 
\caption{{\protect\small Spin and transport gaps as a function of Co
concentration. The Curie-Weiss behaivor is widely identified in the series.
Both spin and transport gaps are shrinking due to Co impurity states, which
likely reside in the gap.}}
\label{Fig4}
\end{figure}
%

\subsection{Electronic Transport}

Whereas the electronic transport of FeSb$_{2}$ is strongly anisotropic,
there is much less dependence on "direction" for Fe$_{1-x}$Co$_{x}$Sb$_{2}$.
Figure 3 shows the normalized resistivity as a function of temperature along
all three axes. Absolute values of the resistivity at $T=350$ $K$ for
samples with doping up to $x=0.4$ are listed in Tab. II. For $x=0$ (FeSb$%
_{2} $), there is metallic behavior above 40 K along the \b{a}xis, and
activated behavior below that temperature. On the other hand, FeSb$_{2}$ is
semiconducting for current applied along \a and \c{a}xes. With a small
amount of doping ($x=0.1$), the resistivity exhibits good metallic behavior
over the whole temperature range, and showing very small defect scattering ($%
\rho _{0}\sim 10\,\mu \Omega \mathrm{cm}$) at 1.8 K, the lowest measured
temperature.

For $0.1<x<0.4$ the electronic transport is metallic for current applied
along all three principal crystalline axes. Metallicity induced by Co
substitution is suppressed as the system evolves towards the region of the
structural phase transition for $x\geq 0.5$, due to monoclinic distortion
that causes doubling of the \c{a}xis and the consequent reduction of the
overlap of d$_{xy}$ orbitals. For $0.5\leq x\leq 1\,$($P21/c$ structure) a
small anisotropy appears again, but it has different character comparing it
to $0\leq x<0.5$ ($Pnnm$ structure). The structural change is likely one
origin of that difference. More theoretical and experimental band structure
work would be useful to shed light on the anisotropy of electronic transport
of Fe$_{1-x}$Co$_{x}$Sb$_{2}$($0.5\leq x\leq 1\,$).

For the semiconducting samples, we identify a regime of thermally activated
resistivity above 40 K described by 
\begin{equation*}
\rho =\rho _{0}e^{\frac{{\Delta _{\rho }}}{{2k_{B}T}}}.
\end{equation*}%
Here $\Delta _{\rho }$ is the transport energy gap. The activation energy
for the transport gap is obtained at temperatures above 40 K. The spin gaps
are obtained from the narrow-band-small-gap model of the polycrystalline
magnetic susceptibility. For x=0.05 the average value of $\Delta _{\rho }$
from $\rho _{a}(T)$ and $\rho _{c}(T)$ is shown(they agree within
measurement error$\sim 10\%$), since $\rho _{b}$ is metallic. For $0.5\leq
x<1$ the value of $\Delta _{\rho }$ from $\rho _{c}$ is shown since $\rho
_{a}$ and $\rho _{b}$ are metallic. Near structural phase transition, the
value of $\Delta _{\rho }$ from $\rho _{a}$ is shown since $\rho _{b}$ and $%
\rho _{c}$ are metallic. A discrepancy between the spin and the transport
gap values is evident. A possible difference between the gaps in charge and
spin excitation channels has also been observed in pure FeSb$_{2}^{\text{ \ }%
9}$ as well as in FeSi$^{16,17}$.

\begin{table}[h]
\caption{Values of resistivity Fe$_{1-x}$Co$_{x}$Sb$_{2}$(0$\leq $x$\leq $%
0.4) at 350 K.}
\label{Tb2}
\begin{center}
\begin{tabular}{m{0.76in}m{0.76in}m{0.76in}m{0.76in}}
\hline\hline
$x$ & $\rho _{a}(m\Omega \mathrm{cm})$ & $\rho _{b}(m\Omega \mathrm{cm})$ & $%
\rho _{c}(m\Omega \mathrm{cm})$ \\ \hline
$0$ & $0.79$ & $0.27$ & $0.63$ \\ 
$0.05$ & $0.74$ & $0.23$ & $0.26$ \\ 
$0.1$ & $0.39$ & $0.20$ & $0.24$ \\ 
$0.2$ & $0.96$ & $0.59$ & $0.68$ \\ 
$0.3$ & $1.38$ & $0.50$ & $0.69$ \\ 
$0.4$ & $1.80$ & $1.04$ & $1.56$ \\ \hline\hline
\end{tabular}%
\end{center}
\end{table}

In FeSb$_{2}$ application of a magnetic field induces a large anisotropic
positive magnetoresistance for current applied along the highly conductive $%
\hat{b}$ axis.$^{9}$ This effect is strongly enhanced in Fe$_{1-x}$Co$_{x}$Sb%
$_{2}$, particularly for highly metallic samples. As an example, in Fig. 6
the effect in Fe$_{0.9}$Co$_{0.1}$Sb$_{2}$ is shown. In addition,
magnetoresistance for pure FeSb$_{2}$ is plotted for comparison. Whereas
magnetic field has little effect on the semiconducting electronic transport
along the \a and $\hat{c}$ axes in FeSb$_{2}$, for the $\hat{b}$ axis
transport we observe up to two orders of magnitude of positive
magnetoresistance at temperatures in the vicinity of the metal to
semiconductor crossover temperature. Away from the crossover temperature the
effect diminishes, below 5 K and in the vicinity of room temperature the
application of a magnetic field has no effect on the $\hat{b}$ axis
transport.

%
\begin{figure}[b]
\centerline{\includegraphics[scale=0.66]{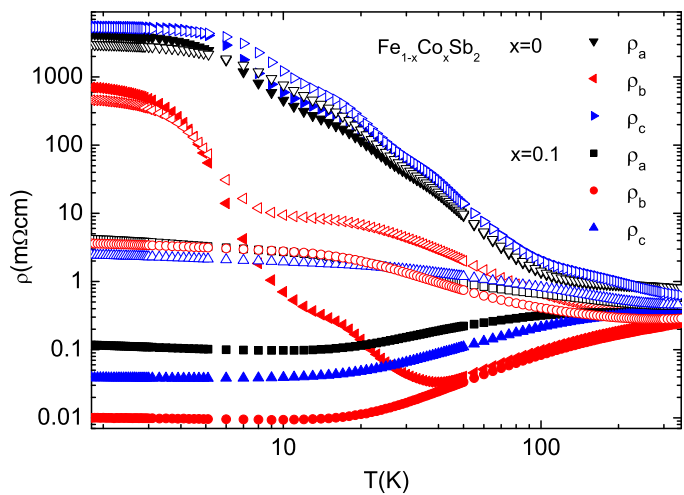}} 
\vspace*{-0.2cm}
\caption{{\protect\small Large magnetoresistance for $x=0$ and moderately
doped ($x=0.1$) highly metallic samples. Solid and open symbols denote data
in 0 and 90 kOe applied fields, respectively.}}
\label{Fig5}
\end{figure}
%

For $x=0.1$, this strong effect is more isotropic and is observed for
current applied along all three crystalline axes. However for $I\,||\,\hat{b}
$, the effect is strongest. Below 10 K, the resistivity increases up to 2.5
orders of magnitude in an applied field of $H=90$ kOe. The transverse
magnetoresistance (MR), defined as 
\begin{equation*}
\mathrm{MR}={\frac{{[\rho (90\,\mathrm{kOe})-\rho (0)]}}{{\rho (0)}},}
\end{equation*}%
reaches $36000\,\%$ at $T=1.8$ K and $\sim $ 30 \thinspace \% at $T=300$ K
in 90kOe. This effect is comparable to the colossal magnetoresistance
observed below room temperature in manganites. Furthermore, it is much
higher at room temperature than in manganites. The origin of the
magnetoresistance in FeSb$_{2}$(x=0) is most likely due to the large
mobility ($\mu $) of carriers. The mobile carriers in a magnetic field move
in cyclotron orbits specified by $\omega _{c}\tau \sim \mu H>1$. Therefore,
a strong positive magnetoresistance effect is expected in even a moderate
field.$^{9}$

Magnetic isotherms deviate from the quadratic relation for the ordinary
metallic case, i.e. $\Delta \rho (H)/\rho _{0}\propto H^{2}$ and there is a
poor Kohler scaling for Fe$_{0.9}$Co$_{0.1}$Sb$_{2}$ . The spin fluctuation
mechanism proposed by Y. Takahashi and T. Moriya $^{3}$ can be discarded
based on the sign of the effect. Mainly because the spin fluctuation theory
predicts negative magnetoresistance due to the thermally suppressed spin
fluctuation amplitude. In addition, the observed temperature dependence of
magnetization does not contain $T^{3/2}$ and $T^{5/2}$ terms that arise from
spin wave and spin wave - spin wave interactions.


\subsection{Heat Capacity}

The heat capacity $C_{p}/T$ as a function of T$^{2}$ below 10 K is shown in
Fig. 6. We observe a weak but clear signature of a ferromagnetic transition
in the vicinity of the $T_{C}$, as derived from magnetization measurements. 
\begin{figure}[b]
\centerline{\includegraphics[scale=0.65]{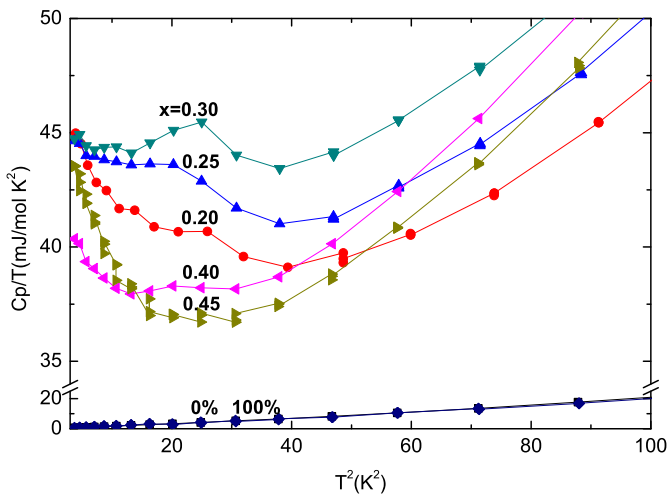}} 
\vspace*{-0.2cm}
\caption{{\protect\small C$_{p}/T$ as a function of $T^{2}$. Cusps
correspond to ferromagnetic transition. Co doping gives rise to substantial
increase in gamma. }}
\label{Fig6}
\end{figure}
%
%
By integrating $C_{p}/T$ over $T$ under this anomaly, the magnetic entropy
contribution per Co is evaluated. For $x=0.3$, we find $\Delta
S=0.35\%\,k_{B}\ln 2$, in good agreement with the results obtained from the
magnetization measurement. The end members of the Fe$_{1-x}$Co$_{x}$Sb$_{2}$
series have negligible electronic specific heat $\gamma =9.2\times 10^{-3}$
mJ/mol K$^{2}$. This implies an negligible density of states at the Fermi
level, i.e. an insulating state. The result agrees well with optical
spectroscopy measurements that show a negligible Drude weight of $\sigma
(\omega )$ at low frequencies.$^{18}$

The changes in $\gamma $ as a function of $x$ are correlated with the
evolution of the magnetic properties, since the saturation moment dependence
for samples in the ferromagnetic region reaches a maximum value for $x=0.3$.
Within the Kondo insulator framework, a metallic state with a large carrier
mass enhancement is generated by a carrier induced closing of the
hybridization gap. Consequently, the carrier mass enhancement leads to large 
$\gamma $ value due to a renormalized band structure. The $\gamma $ values
observed in Fe$_{1-x}$Co$_{x}$Sb$_{2}$ are comparable to those in the
ferromagnetic regime of FeSi$_{1-x}$Ge$_{x}$$^{16}$ and FeSi$_{1-x}$Al$_{x}$%
, Fe$_{1-x}$Co$_{x}$Si as well.$^{17,19}$ To estimate the carrier mass
enhancement, we make the free-electron assumption. That is, 
\begin{equation*}
\gamma =\frac{\pi ^{2}k_{B}^{2}N(E_{F})}{3}=\frac{k_{B}^{2}k_{F}m^{\ast }}{%
3\hbar ^{2}}
\end{equation*}%
with $k_{F}=(3\pi ^{2}n)^{1/3}$. The carrier density is denoted by $n$ and
is estimated to be equal to the Co concentration assuming one itinerant
electron per Co$^{2+}$. This model implies only a small enhancement of
carrier mass and disagrees with the Kondo Insulator model. However, Hall
constant measurements would be useful to determine the precise carrier
density and so provide a definitive test for the applicability of the Kondo
Insulator model. Furthermore, these measurements will help shed more light
on the origin of the colossal magnetoresistance in this system.

\begin{table}[t]
\caption{Sommerfeld coefficient and effecitve mass of carriers.}
\label{Tb3}%
\begin{tabular}{cp{0.5in}m{1.08in}m{0.5in}c}
\hline\hline
${\large x}$ &  & ${\large \gamma \mathrm{(mJ/molK}}^{2}{\large )}$ &  & $%
{\large m}^{\ast }{\large (m}_{e}{\large )}$ \\ \hline
\multicolumn{1}{c}{${\large 0.20}$} &  & \multicolumn{1}{c}{${\large 25.9}$}
&  & \multicolumn{1}{c}{${\large 1.35}$} \\ 
\multicolumn{1}{c}{${\large 0.25}$} &  & \multicolumn{1}{c}{${\large 27.6}$}
&  & \multicolumn{1}{c}{${\large 1.33}$} \\ 
\multicolumn{1}{c}{${\large 0.30}$} &  & \multicolumn{1}{c}{${\large 29.96}$}
&  & \multicolumn{1}{c}{${\large 1.33}$} \\ 
\multicolumn{1}{c}{${\large 0.40}$} &  & \multicolumn{1}{c}{${\large 22.9}$}
&  & \multicolumn{1}{c}{${\large 0.95}$} \\ 
\multicolumn{1}{c}{${\large 0.45}$} &  & \multicolumn{1}{c}{${\large 22.4}$}
&  & \multicolumn{1}{c}{${\large 0.89}$} \\ \hline\hline
\end{tabular}%
\end{table}

\section{Conclusion}

In summary, the Fe$_{1-x}$Co$_{x}$Sb$_{2}$ electronic system exhibits two
consecutive metal to semiconductor crossovers. The first is caused by
electron doping of the semiconducting ground state of FeSb$_{2}$ for $x=0.1$%
. The second is induced by a structural change for doping levels in the
vicinity of $x=0.5$. The density of states at the Fermi level increases due
to the increased Co concentration, as shown in the rise of the electronic
specific heat. The weak ferromagnetism induced by the polarization of a
small portion of Co$^{2+}$ ions arises in the metallic state for $0.2\leq
x<0.5$. A colossal magnetoresistance as large as 36000\% is observed in the
vicinity of the first metal to semiconductor crossover. The overall
properties of Fe$_{1-x}$Co$_{x}$Sb$_{2}$ can be understood in the
narrow-band-small-energy-gap framework where Co substitution gives rise to
impurity states in the gap, metallicity, and Curie-Weiss behavior. This is
further supported by the recent LDA+U calculation of Lukoyanov on FeSb$_{2}$%
, which indicates sharp peaks in the DOS at the edge of a small gap.$^{20}$ Possible consequence of the presence of magnetic Fe or an unknown ferromagnetic impurity would imply that the value of the direct Coulomb repulsion parameter U in ref. 20 does not exceed the critical value of U$_{C}$=2.6eV for all values of x in Fe1-xCoxSb2 and that the weak ferromagnetism is extrinsic.$^{20}$

\section{Acknowledgments}

We thank Donavan Hall, T. M. Rice, P. C. Canfield, S. L. Bud'ko, Myron
Strongin and M. O. Dzero for useful communication and John Warren for SEM
measurement. This work was carried out at the Brookhaven National
Laboratory, which is operated for the U.S. Department of Energy by
Brookhaven Science Associates (DE-Ac02-98CH10886). This work was supported
by the Office of Basic Energy Sciences of the U.S. Department of Energy.


\begin{thebibliography}{99}
\bibitem{Foex} G. F\"{o}ex, J. Phys. Radium \textbf{9}, 37 (1938).

\bibitem{Jaccarino} V. Jaccarino, G. K. Wertheim, J. H. Wernick, L. R.
Walker, and S. Arajs, Phys. Rev. \textbf{160}, 476 (1967).

\bibitem{Moriya} Y. Takahashi and T. Moriya, J. Phys. Soc. Jpn. \textbf{46}%
,1451 (1979).

\bibitem{Fisk1} G. Aeppli and Z. Fisk, Comments Condens. Matter Phys. 
\textbf{16}, 155 (1992).

\bibitem{Varma} C. M. Varma, Phys. Rev. B \textbf{50}, 9952 (1994).

\bibitem{Rice} V. I. Anisimov, S. Y. Ezhov, I. S. Elfimov, I. V. Solovyev
and T. M. Rice, Phys. Rev. Lett. \textbf{76}, 1735 (1996).

\bibitem{Sarrao} D.Mandrus, J.L.Sarrao, A.Migliori, J.D.Thompson and Z.Fisk,
Phys. Rev. B \textbf{51}, 4763 (1995).

\bibitem{Park} C.-H.Park, Z.-X.Shen,A.G.Loeser and D.S.Dessau, Phys. Rev. B 
\textbf{52}, 16981 (1995)

\bibitem{Canfield1} C. Petrovic, J. W. Kim, S. L. Bud'ko, A. I. Goldman and
P. C. Canfield, Phys. Rev. B \textbf{67}, 155205 (2003).

\bibitem{Budko} C. Petrovic, Y. Lee, T. Vogt, N. Dj. Lazarov, S. L. Bud'ko
and P. C. Canfield, Phys. Rev. B \textbf{72}, 045103 (2005).

\bibitem{Canfield} P. C. Canfield, Z. Fisk Phil. Magaz. B \textbf{65}, 1117
(1992).

\bibitem{Fisk} Z. Fisk, J. P. Remeika, in: K. A. Gschneider, J. Eyring
(Eds.), Handbook on the Physics and Chemistry of Rare Earths, \textbf{Vol. 12%
}, Elsevier, Amsterdam, (1989).

\bibitem{Hunter} Hunter B., "Rietica - A visual Rietveld program",
International Union of Crystallography Commission on Powder Diffraction
Newsletter No. \textbf{20}, (Summer) http://www.rietica.org (1998)

\bibitem{Einar} Einar Bjerkelund and Arne Kjekshus, Acta Chem. Scand. 
\textbf{24},3317-3325(1970).

\bibitem{Goodenough} J. B. Goodenough, J. Solid State Chem. \textbf{5}, 144
(1972).

\bibitem{Kennedy} S. Yeo, S. Nakatsuji, A. D. Bianchi, P. Schlottmann, Z.
Fisk, L. Balicas, P. A. Stampe, and R. J. Kennedy, Phys. Rev. Lett. \textbf{%
91}, 046401 (2003).

\bibitem{Ramirez} J.F.DiTusa, K.Friemelt, E.Bucher, G.Aeppli, A.P.Ramirez,
Phys. Rev. B \textbf{58}, 10288 (1998).

\bibitem{Perucchi} A. Perucchi, L. Degiorgi, R. Hu, C. Petrovic, and V.
Mitrovi{\'c}, Eur. Phys. J. B \textbf{54}, 175 (2006).

\bibitem{Mandrus} M. A. Chernikov, L. Degiorgi, E. Felder, S. Paschen, A. D.
Bianchi, H. R. Ott, J. L. Sarrao, Z. Fisk, and D. Mandrus, Phys. Rev. B 
\textbf{56}, 1366 (1997).

\bibitem{TRice} A. V. Lukoyanov, V. V. Mazurenko, V. I. Anisimov, M.Sigrist,
and T. M. Rice, Eur. Phys. J. B \textbf{53}, 205 (2006).
\end{thebibliography}
\end{document}